\documentclass[aps,prl,twocolumn,preprintnumbers,amsmath,superscriptaddress,amssymb]{revtex4-1}
\usepackage{graphicx}
\usepackage{tabularx}
\usepackage{euscript}
\usepackage{epsfig,psfrag,subfigure}
\usepackage{dcolumn}
\usepackage{exscale}
\usepackage{amsbsy}
\usepackage{bm}
\usepackage{color}
\usepackage{amsfonts}

\def\avg#1{\left\langle#1\right\rangle}

\def\abs#1{\left|#1\right|}

\def\be{\begin{equation}}       \def\ee{\end{equation}}
\def\bea{\begin{eqnarray}}      \def\eea{\end{eqnarray}}
\def\ba{\begin{array} }
\def\ea{\end{array} }
\def\bnum{\begin{enumerate} }
\def\enum{\end{enumerate}}

\def\nn{\nonumber}

\def\=>{\Rightarrow}
\def\>{\rightarrow}
\def\A{\uparrow}
\def\V{\downarrow}

\def\eye2{Fathbb{I}}

\def\Eq#1{Eq.~(\ref{#1})}
\def\Fig#1{Fig.~\ref{#1}}

\renewcommand{\v}[1]{{\bf #1}}

\newcommand{\s}{{\sigma}}

\renewcommand{\>}{\rangle}

\newcommand{\e}{\epsilon}

\begin{document}

\title{Fractional charge and emergent mass hierarchy in diagonal two-leg $t$-$J$ cylinders}
\author{Yi-Fan Jiang}
\affiliation{Institute for Advanced Study, Tsinghua University, Beijing 100084, China}
\affiliation{Department of Physics, Stanford University, Stanford, California 94305, USA}
\author{Hong-Chen Jiang}
\affiliation{Stanford Institute for Materials and Energy Sciences, SLAC National Accelerator Laboratory and Stanford University, Menlo Park, CA 94025, USA}
\author{Hong Yao}
\affiliation{Institute for Advanced Study, Tsinghua University, Beijing 100084, China}
\affiliation{Collaborative Innovation Center of Quantum Matter, Beijing 100084, China}
\author{Steven A. Kivelson}
\affiliation{Department of Physics, Stanford University, Stanford, California 94305, USA}
\begin{abstract}
We define a new class of ``diagonal'' $t$-$J$ ladders rotated by $\pi/4$ relative to the canonical lattice directions of the square lattice, and study it using  density matrix renormalization group (DMRG). Here, we focus on the two-leg cylinder with a doped hole concentration near $x=1/4$. At exactly $x=1/4$, the system forms a period 4 charge density wave (CDW) and exhibits spin-charge separation. Slightly away from $1/4$ doping we observe several topologically distinct types of solitons with well defined fractionalized quantum numbers. Remarkably, given the absence of any obvious small parameter, the effective masses of the  emergent solitons differ by several orders of magnitude.
\end{abstract}
\date{\today}
\maketitle

{\bf Introduction:} As a paradigm for the description of high temperature superconductors\cite{FCZhang88}, the $t$-$J$ model, and the closely related Hubbard model have been studied extensively by many different numerical methods and are thought to possess a rich phase diagram\cite{Ogata91,Dagotto92,Dagotto94,White98,White99,Himeda02, White04, White09, Corboz11, Liu12, Gu13, Corboz14, unpublished}. In most of these studies, the system is taken to be oriented parallel to the primitive lattice vectors of the square lattice. However, in attempting to extrapolate the results to the thermodynamic limit in 2D, it is also useful to study ladders  with different geometries\cite{Fouet06,Berg09,LeBlanc15}.

\begin{figure}[t]
\centering
\includegraphics[width=7cm]{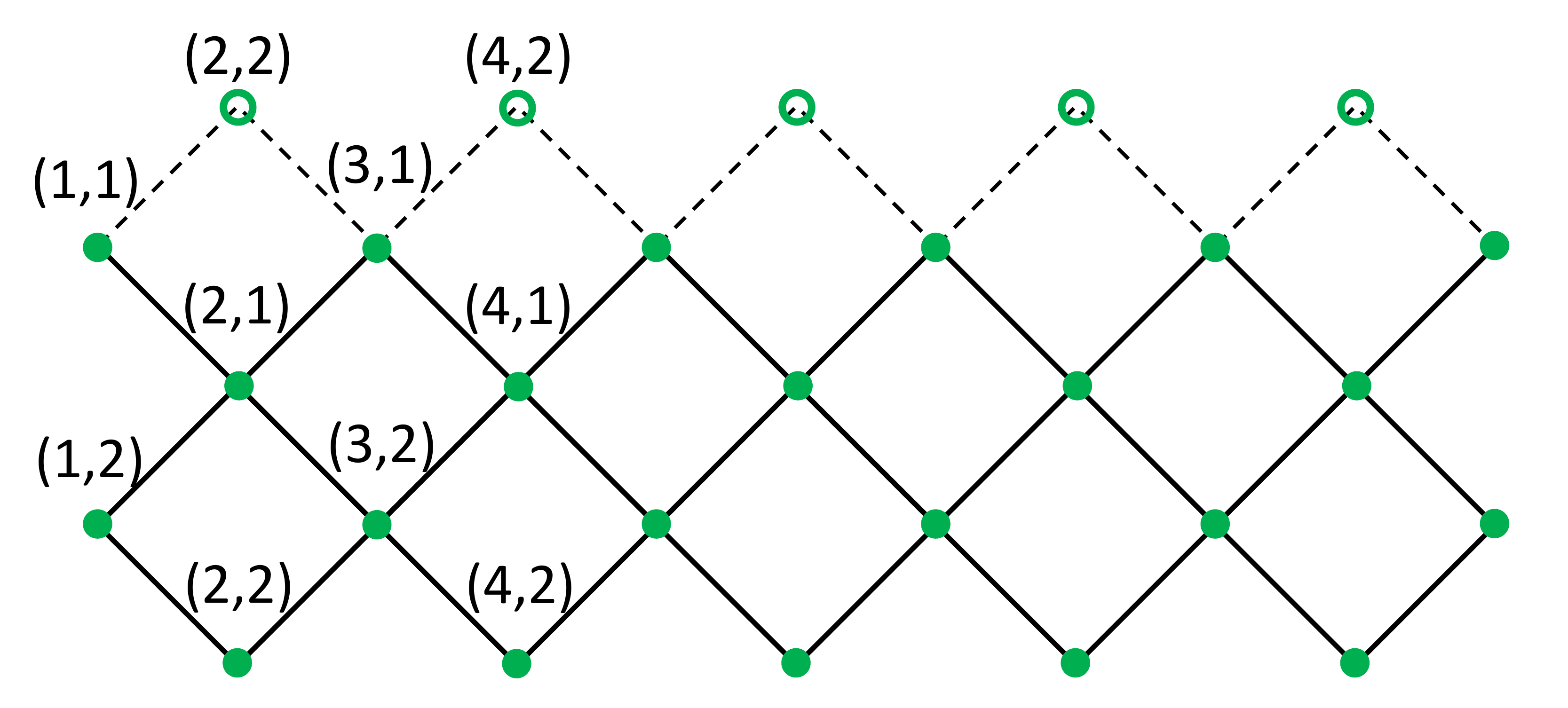}
\caption{Diagonal two-leg ladder with cylinder boundary condition (CBC); empty circles and dashed lines represent the periodic boundary.}
\label{twoleg}
\end{figure}

A diagonal cylinder, rotated by $\pi/4$ relative to the primitive lattice directions of the sort shown in \Fig{twoleg}, has several advantages over the usual one. For example, since a mirror symmetry along the unit cell diagonal is preserved, it is possible to make sharp distinctions between states whose signatures on a regular ladder would be identical -- for instance, one can distinguish $d$-wave superconductivity from $s$-wave superconductivity and vertical ``stripe'' (unidirectional CDW) order from ``checkerboard'' (bidirectional CDW) order, and a nematic phase would correspond to a phase that spontaneously breaks this mirror symmetry.  Moreover, while on usual ladders of width larger than 2-legs, there is a clear tendency for stripe order to come at the expense of long-range superconducting coherence, on diagonal cylinders of appropriate width, CDW order resembling the stripes on a barber pole can involve infinite length stripes, which might therefore compete less strongly with superconducting coherence.

Here we present the first results of a planned extensive DMRG\cite{White92,White93} study of the $t$-$J$ model on diagonal ladders. Although our principle interest in this model concerns the extrapolation to 2D, it is also of interest in the context of multi-component 1D systems.  Indeed, the results concerning the 2-leg cylinder near $x=1/4$ doping are already interesting from this 1D perspective.

\begin{table}[b]
\includegraphics[height=2.1cm]{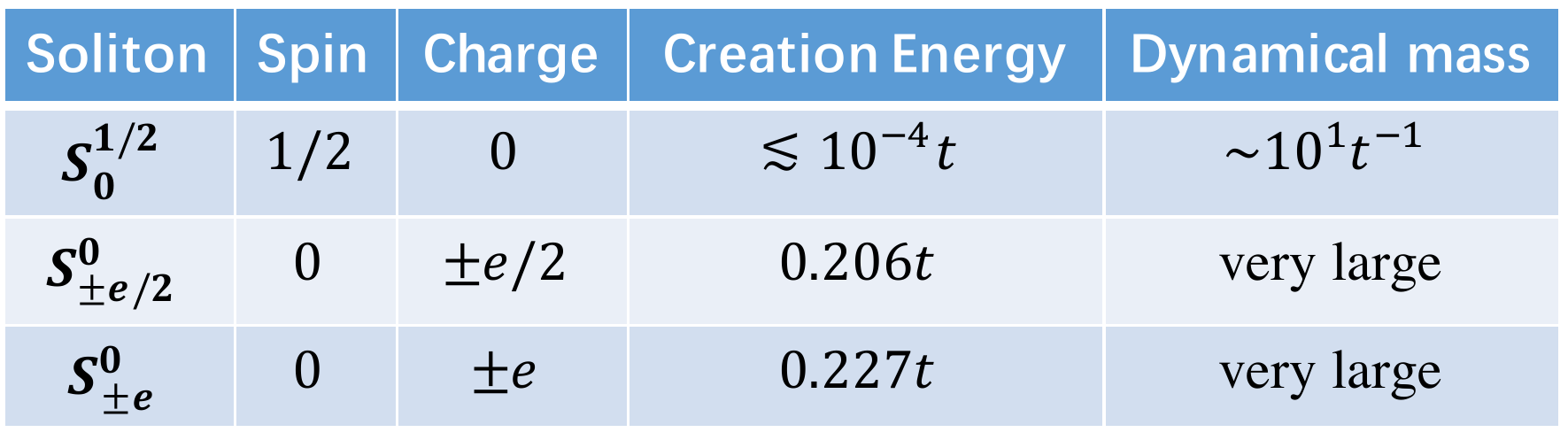}
\caption{Physical quantities of three kinds of solitons. The solitons are illustrated in \Fig{soliton}. For charged solitons, their creation energies refer to half of the energy cost of creating a pair of solitons with opposite charge. The dynamical mass is defined as the zero-point energy to confine a soliton to a region of size $L$ according to $E \sim \frac{1}{2 M^*}(\frac{\pi}{L})^2$.}
\label{table1}
\end{table}

At precisely $x=1/4$, the system exhibits an interesting commensurate CDW with long-range order.  While the period of the density wave order is 4 lattice constants, the period 2 ``harmonic'' is highly dominant and the period 4 ``fundamental'' is extremely weak. Looking at the excitation spectrum, in contrast to the usual 2-leg ladder, this diagonal ladder exhibits clear spin-charge separation. Indeed, multiple types of fractionalized soliton excitations arise with different topological characters and associated with different (fractional) quantum numbers, as presented in Table \ref{table1}. These solitons are somewhat analogous to the solitons that arise in the mean-field solution of the electron-phonon (commensurate Peierls) problem\cite{Zabusky65,Zhang87,Heeger88}, but here they arise directly from the strong electronic correlations. In particular, we identify two flavors of solitons -- one is a highly-local charge excitation with a large creation energy, while the other is an extended spin excitation (spinon) with a creation energy that is at least several orders smaller.

{\bf Model:} The Hamiltonian we study in this paper is the nearest neighbor $t$-$J$ model:
\be
H=-t\sum_{\avg{ij}\s}(c^\dagger_{i\s}c_{,j\s}+h.c.)+J\sum_{\avg{ij}}(\v{S}_i\cdot\v{S}_j-\frac14n_i n_j),
\ee
where $t>0$ is the uniform hopping integral, $J>0$ is the superexchange coupling, $c_{i\s}$ is the electron annihilation operator at site $i=(x,y)$ with spin polarization $\s=\uparrow$/$\downarrow$, $\v{S}$ is the spin operator, $n_i=\sum_{\s}c^{\dagger}_{i\s}c_{i\s}$ is the  electron density, and $\avg{ij}$ denotes pairs of nearest neighbor sites.
We henceforth take units of energy such that $t=1$. The Hilbert space has a no-double-occupancy constraint, i.e. $n_i= 0$, 1. The lattice structure of the diagonal two-leg cylinder is illustrated schematically in \Fig{twoleg}. For convenience, we label each site by its location $(x,y)$, where $y$ ranges from 1 to 2 designating the legs and $x$ from 1 to $L$ denoting the position of rungs. In our DMRG simulations of this model, we keep up to 3000 states in the DMRG block and sweep around 30 times such that the truncation error $\e_{trun}$ is at most $10^{-7}$.

\begin{figure}[t]
\centering
\subfigure{\includegraphics[width=4cm]{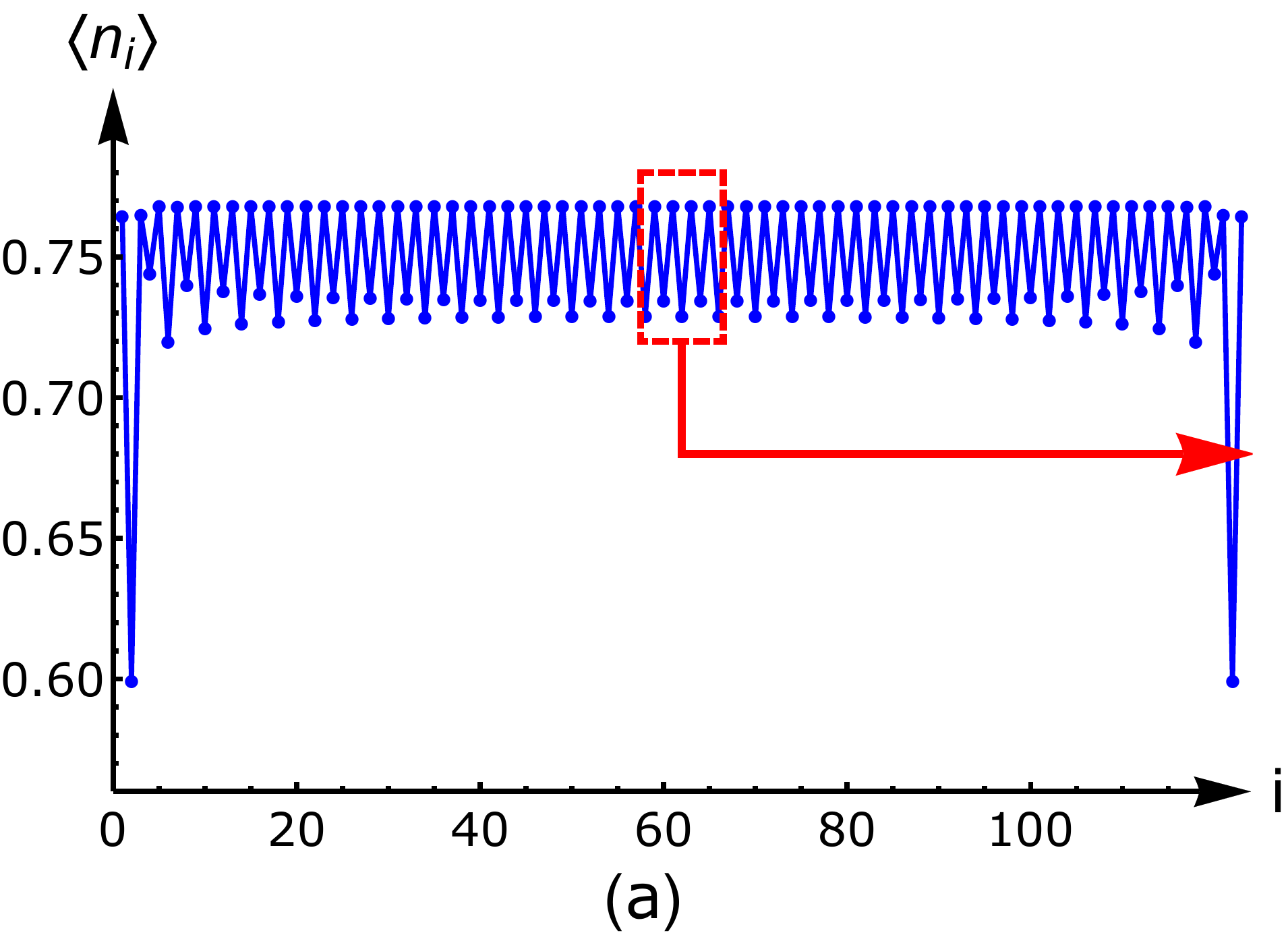}}
\subfigure{\includegraphics[width=4cm]{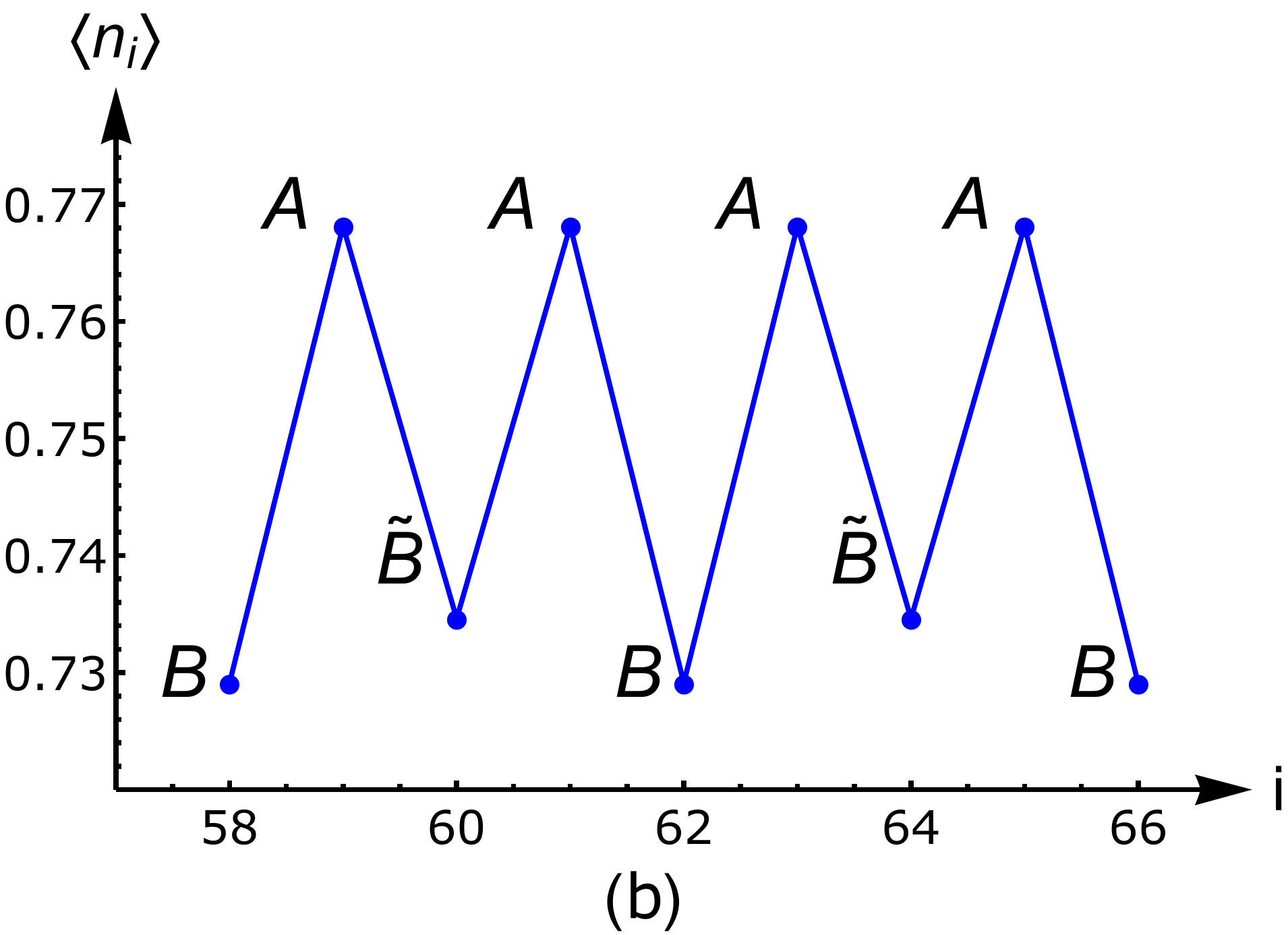}}
\subfigure{\includegraphics[width=4cm]{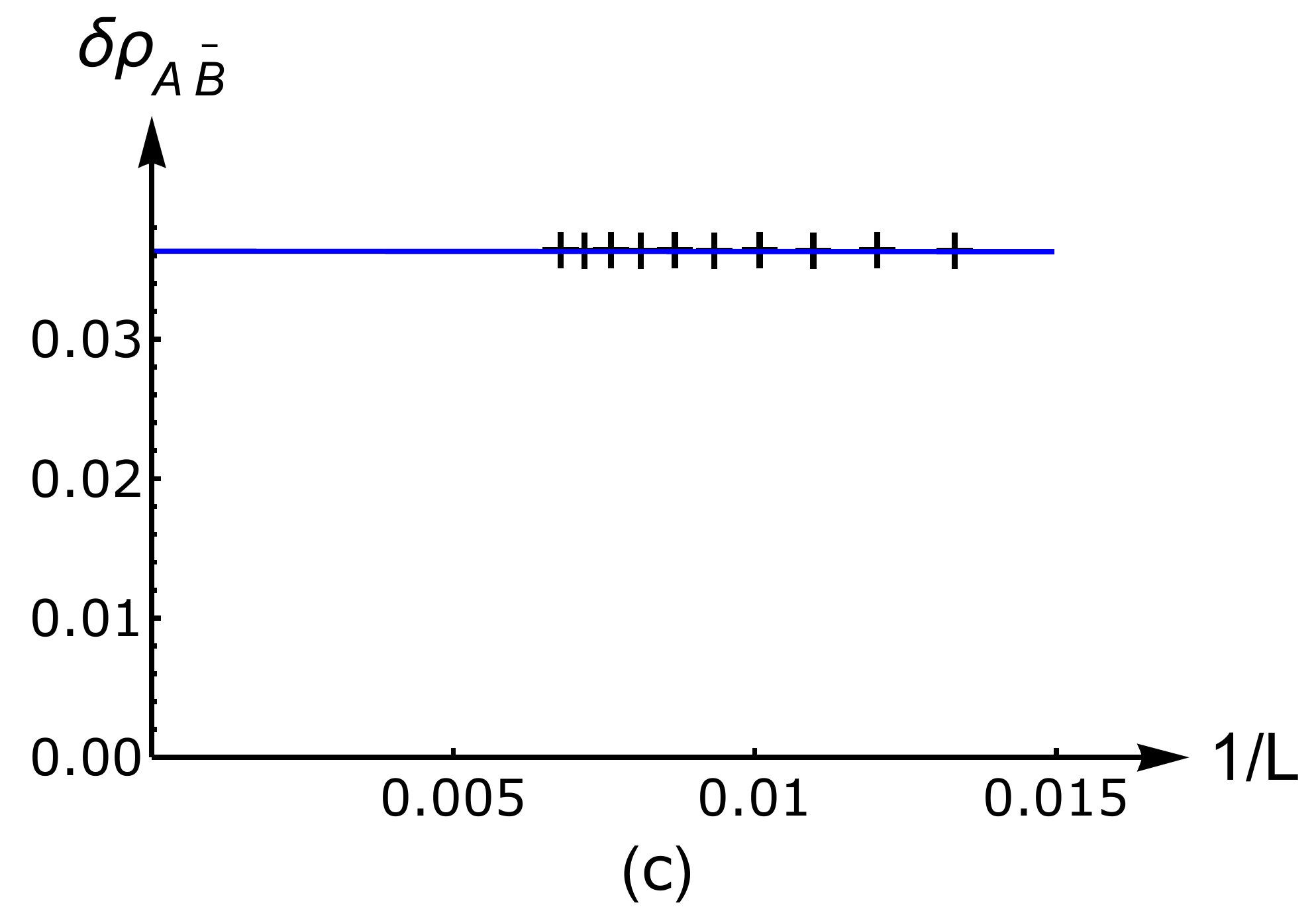}}
\subfigure{\includegraphics[width=4cm]{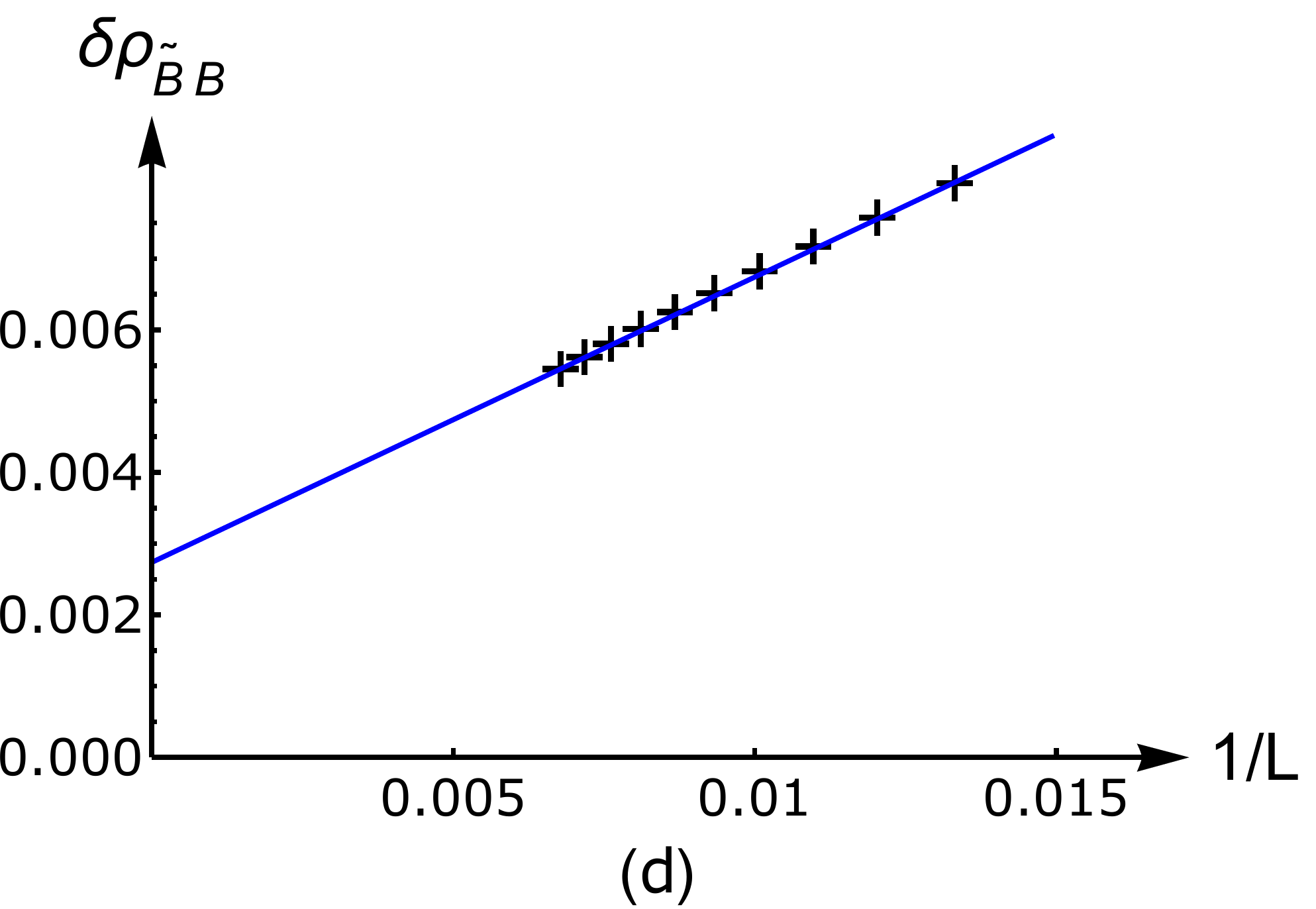}}
\caption{(a) Density profile of the ground state of a $2\times 123$ diagonal cylinder with $2\times 31$ doped holes. We use an odd length cylinder to minimize the boundary effects. (b) Enlarged view of the red rectangle part in (a). There is a period 4 density pattern $ABA\tilde{B}$. (c) The length dependence of $\delta\rho_{A\bar B}\equiv \rho(A)-[\rho(B)+\rho(\tilde B)]/2$; extrapolated to the limit $L\to \infty$ this difference approaches $3.631(3) \times 10^{-2}$. (d) The length dependence of $\delta\rho_{\tilde B B}\equiv \rho(\tilde B)-\rho(B)$; extrapolated to the limit $L\to \infty$ this difference approaches $0.279(3)\times 10^{-2}$.  Here $\rho$ is the averaged density of one type of site in the bulk. The lattice length varies from $L=67$ to $123$.}
\label{rou}
\end{figure}

Note that, besides the global symmetries such as the mirror symmetry along diagonal bonds, the Hamiltonian on the diagonal two-leg cylinder exhibits a local symmetry: exchanging the two sites on any rung preserves the Hamiltonian. Since this local site-exchange symmetry on any
rung is equivalent to a $Z_2$ gauge symmetry, the ground states cannot spontaneously break this symmetry due to the Elitzur's theorem\cite{Elitzur75}. Thus, $\avg{\hat{n}_{x,1}}=\avg{\hat{n}_{x,2}}$ and $\langle\hat{c}^\dagger_{x,1} \hat{c}_{x',1}\rangle=\langle\hat{c}^\dagger_{x,2} \hat{c}_{x',2}\rangle=\langle\hat{c}^\dagger_{x,1} \hat{c}_{x',2}\rangle=\langle\hat{c}^\dagger_{x,2} \hat{c}_{x',1}\rangle$ for $x\neq x'$.  Among other things, this precludes the existence of a nematic phase;  this peculiar local symmetry is not a general feature of wider diagonal ladders or cylinders.

{\bf DMRG Results at $x=1/4$:} As usual, the doping level of the system away from the half-filling is defined as $x=1-\frac{1}{N}\sum_{i\s} \langle c^\dagger_{i\s}c_{i\s} \rangle$, where $N=2L$ is total number of sites. We perform large-scale DMRG simulations to study the $t$-$J$ model on the diagonal two-leg cylinder with open boundary conditions along the leg direction, we adopt a canonical value of $J/t=1/3$, and for present purposes we focus on the doping around $x=1/4$. Since all correlation functions on the legs are exactly same we show numerical results only on the leg $y=1$.

It turns out that the diagonal cylinder at finite doping has many delicate metastable states as shown previously\cite{LeBlanc15} such that its ground states and low energy excitations have not previously been obtained. In order to sort out the lowest energy states by DMRG simulation, we employ the strategy of applying appropriate training fields during the calculations whose details are discussed in the Supplement Material.

The ground-state charge density profile of a $2\times 123$ cylinder with $2\times 31$ holes is shown in \Fig{rou}(a). Although the average value of $x$ differs slightly from 1/4, deep in the bulk ({\it i.e.} far from the open boundary) $x=1/4$, as discussed  below.  We find that the ground state of the system exhibits commensurate period 4 CDW, with a periodic pattern of sites of the form $A B A \tilde B$, as clearly shown in the zoomed in region in \Fig{rou}(b). The difference in the density on the $A$ and the average of the $B$ and $\tilde B$ type sites, $\delta\rho_{A\bar B}\equiv \rho(A)-[\rho(B)+\rho(\tilde B)]/2$,  is an order of magnitude larger than the difference between the $B$ and $\tilde B$ type sites, $\delta\rho_{\tilde B B}\equiv \rho(\tilde B)-\rho(B)$. To understand the significance of this, note that in the limit $\delta\rho_{\tilde B B}\to 0$, the CDW would have period 2; in Fourier transform, this means the ``fundamental'' period 4 mode has a small amplitude $\sim \delta\rho_{\tilde B B}$ while the period 2 first harmonic has a large magnitude $\sim \delta\rho_{A \bar B}$. To obtain a quantitative estimate valid in the thermodynamic limit, we compute  $\delta\rho_{A \bar B}$ and $\delta\rho_{\tilde B B}$ for $L=8n+3$ with various $n$ and then plot the results as a function of $1/L$. As shown in \Fig{rou}(c) and (d), both density differences vary linearly with $1/L$ and approach finite values in the thermodynamic limit: $\delta\rho_{A\bar B}\to 3.631(3)\times 10^{-2}$ and $\delta\rho_{\tilde B B}\to 2.79(3)\times 10^{-3}$.

The ground-state always lies in the spin 0 sector.  However, although, as we discuss below, there are theoretical reasons to expect a spin-gap, if such a gap exists it is exceedingly small.

\begin{figure}[t]
\centering
\subfigure{\includegraphics[width=7cm]{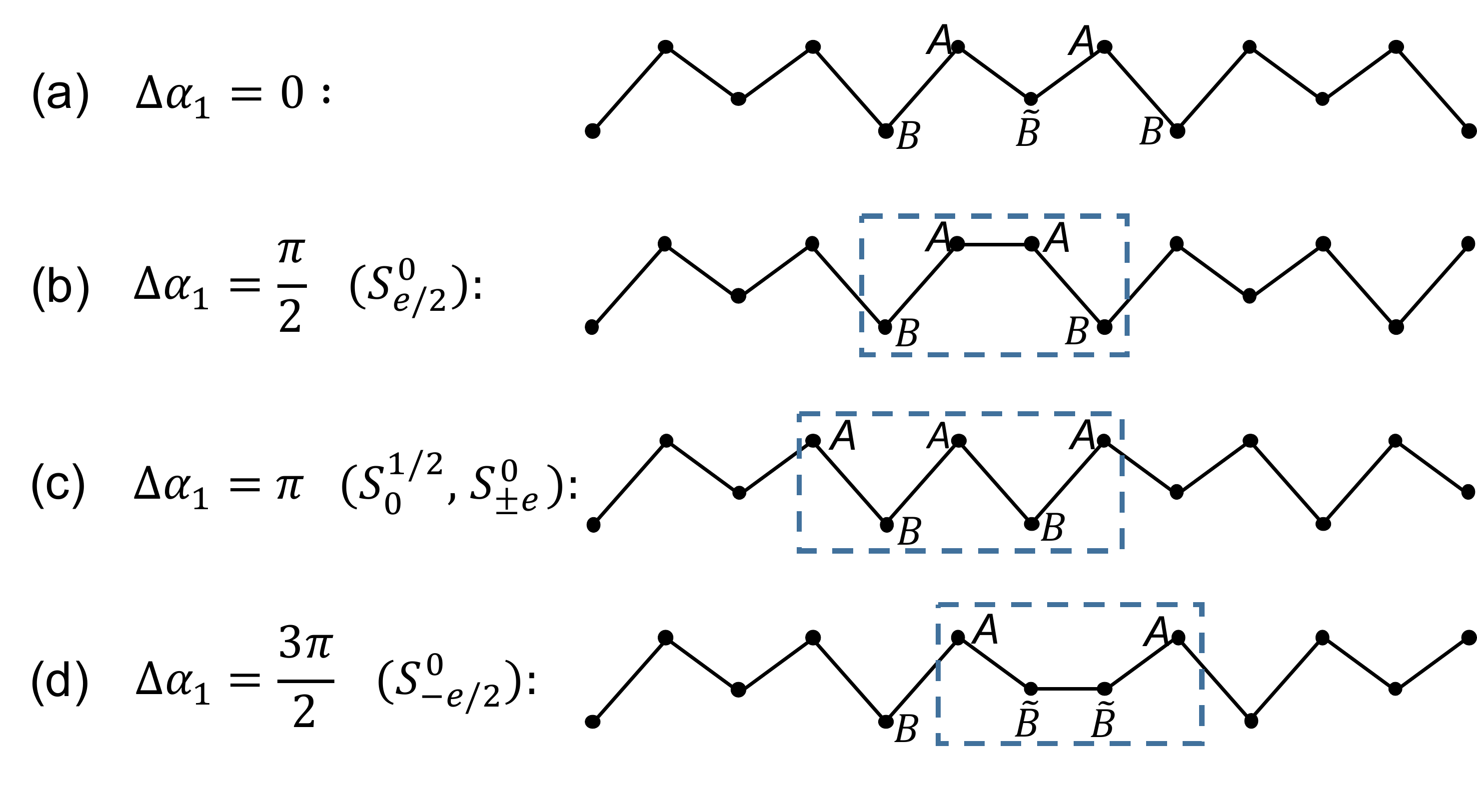}}
\caption{Schematic illustrations of three topologically distinct domain walls. The first chain is a reference without any domain walls. The three dashed rectangles below enclose the domain walls with different subtended angle $\Delta\alpha_1$. Each domain wall is associated with different solitons $S$.}
\label{wall}
\end{figure}

{\bf Solitons in the LG effective field theory:}  As it is an aid to intuition, we can express the CDW in terms of the ground-state configuration of a pair of complex scalar fields, $\phi_1\equiv |\phi_1|e^{i\alpha_1}$ and $\phi_2\equiv |\phi_2|e^{i\alpha_2}$, representing the two harmonics of the density wave:
\be
\rho(x)=\bar{\rho}+\abs{\phi_1}\sin{(\frac{\pi}{2} x+\alpha_1)}
+\abs{\phi_2}\cos{(\pi x+\alpha_2)}
\ee
where $\bar{\rho}=\frac34$ is the average density, and the four symmetry related ground-states correspond to $|\phi_1| =\frac12 \delta\rho_{\tilde B B} \ll |\phi_2| =\frac12 \delta\rho_{A  \bar B}$, $\alpha_2=2\alpha_1$, and $\alpha_1=n\pi/2$ with $n=0$, 1, 2,  and 3. In terms of these fields, we could write an effective Landau-Ginzburg Lagrangian of the form
\bea
\mathcal{L}[\phi_1,\phi_2]&&=\mathcal{L}_1[\phi_1]+\mathcal{L}_2[\phi_2] -\frac{\lambda_1}{4}\big[(\phi_1)^4+c.c.\big]  \\
&&-\frac{\lambda_2}{2}\big[(\phi_2)^2+c.c. \big]
-\lambda_{12} \big[
\phi_2^*(\phi_1)^2 + c.c.\big] +\ldots
\nonumber
\eea
where $\mathcal{L}_j$ are of the usual form as for an incommensurate CDW, and the terms proportional to $\lambda_j$ produce the commensurate lock-in to the lattice.  The term proportional to $\lambda_{12}$ locks the relative phase of the two harmonics, and since it is linear in $\phi_2$, its presence implies that in any state with non-zero $\phi_1$ there will necessarily be an induced (possibly small) harmonic, $\phi_2$. The only really unusual feature here is that the parameters which enter $\mathcal{L}_j$ are such that the ground-state  magnitude of $\phi_2$ is, in fact, much larger than $\phi_1$.

Topological solitons (domain walls) with fractional quantum number appear as  low energy excitations in Peierls systems\cite{Zabusky65,Zhang87,Heeger88}. Analogously, we find stable topological solitons which carry different (fractional) quantum numbers. Specifically, we expect 3 distinct domain walls which can be characterized by the phase change $\Delta \alpha_1$ (subject to the constraint $\Delta\alpha_2=2\Delta\alpha_1$), as shown in Fig. \ref{wall}. From a topological perspective, the $\Delta\alpha_1= \pi$ and $3\pi/2$ domain walls can be viewed as bound-states of, respectively,  two and three $\Delta\alpha_1=\pi/2$ domain walls.

{\bf Solitons from DMRG:}
We induce soliton states by adding  holes or electrons, by flipping spins, or by applying (and then removing) suitable training fields.

\begin{figure}[t]
\centering
\subfigure{\includegraphics[width=4cm]{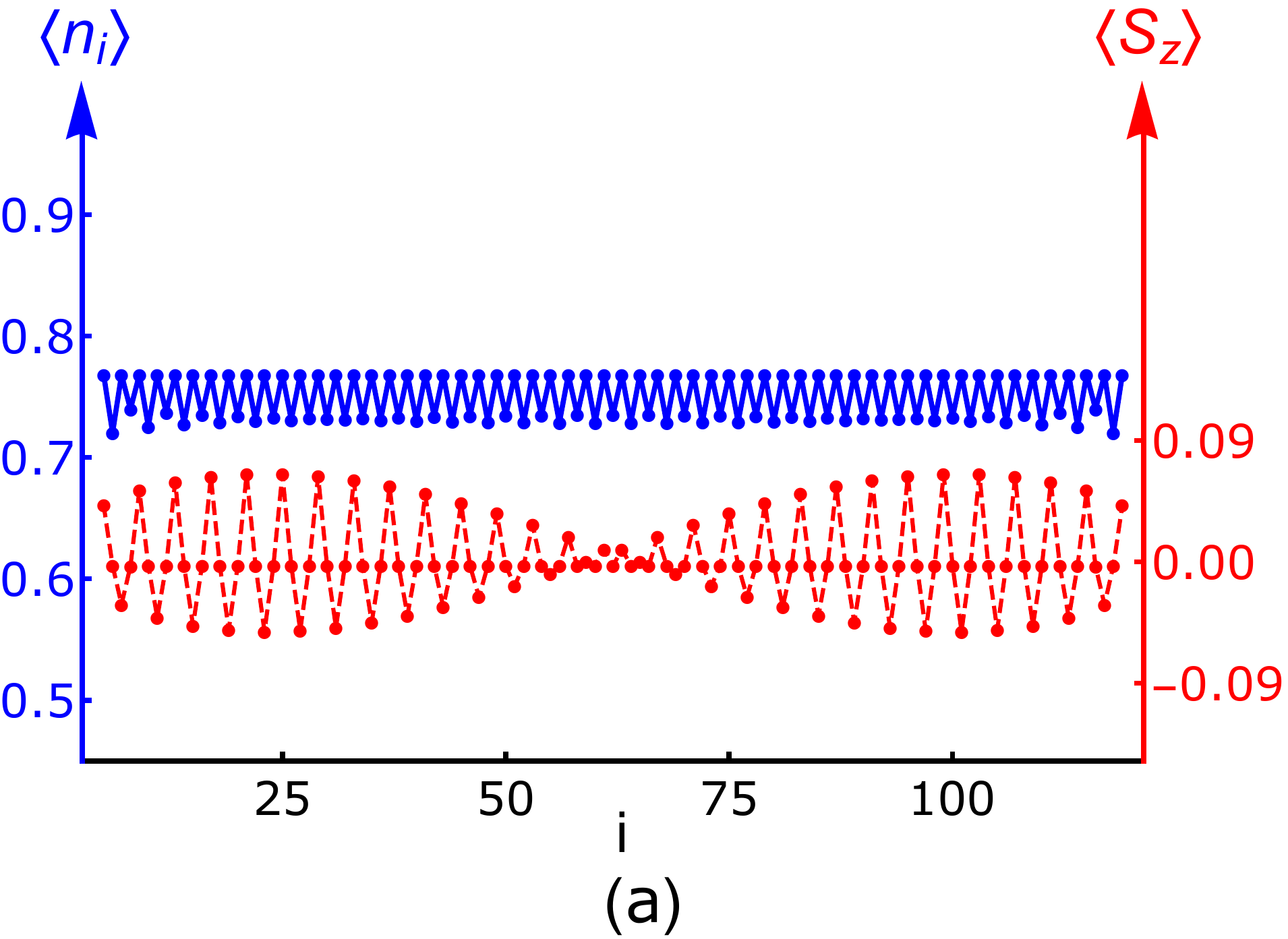}}
\subfigure{\includegraphics[width=4cm]{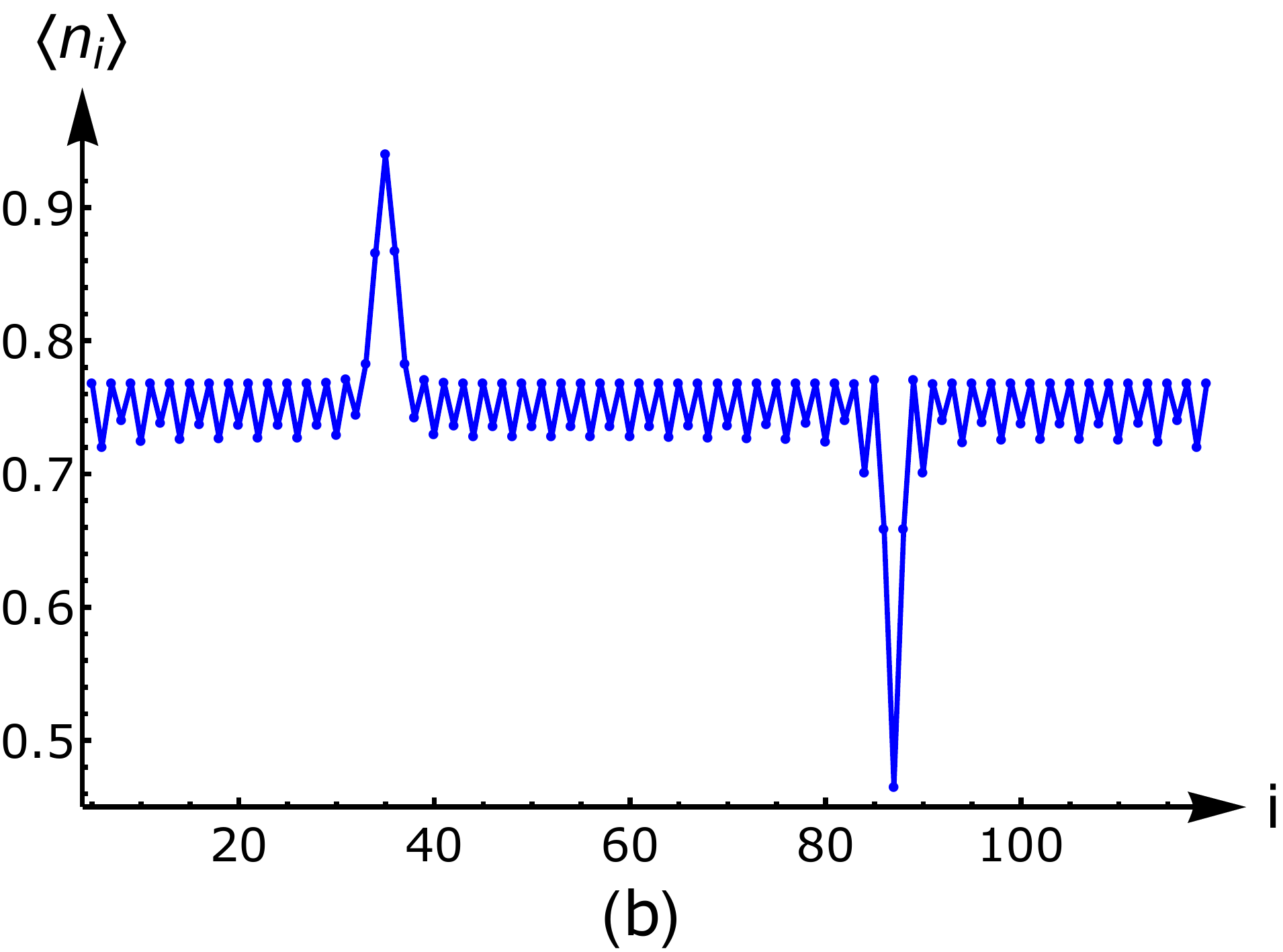}}
\subfigure{\includegraphics[width=4cm]{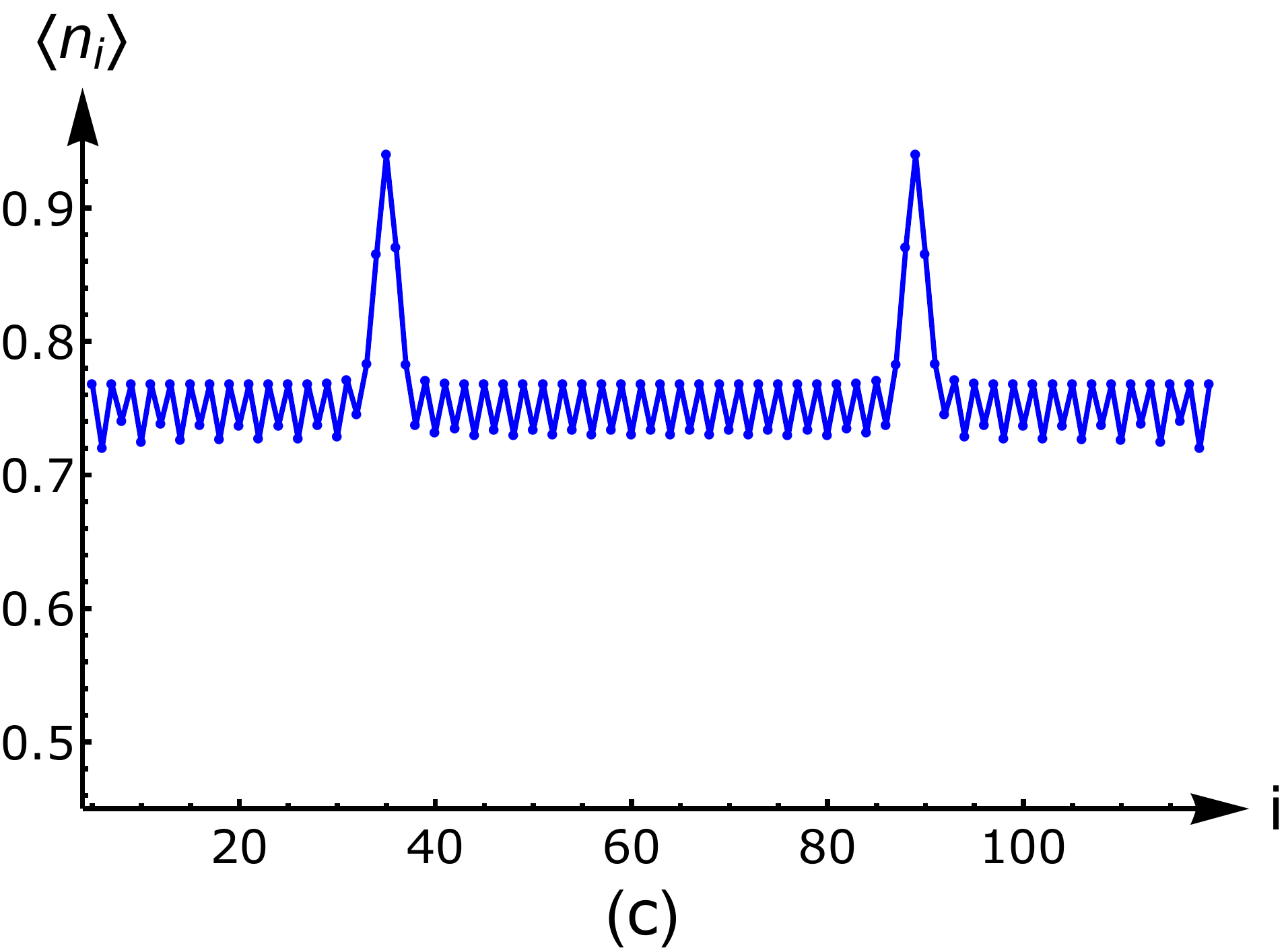}}
\subfigure{\includegraphics[width=4cm]{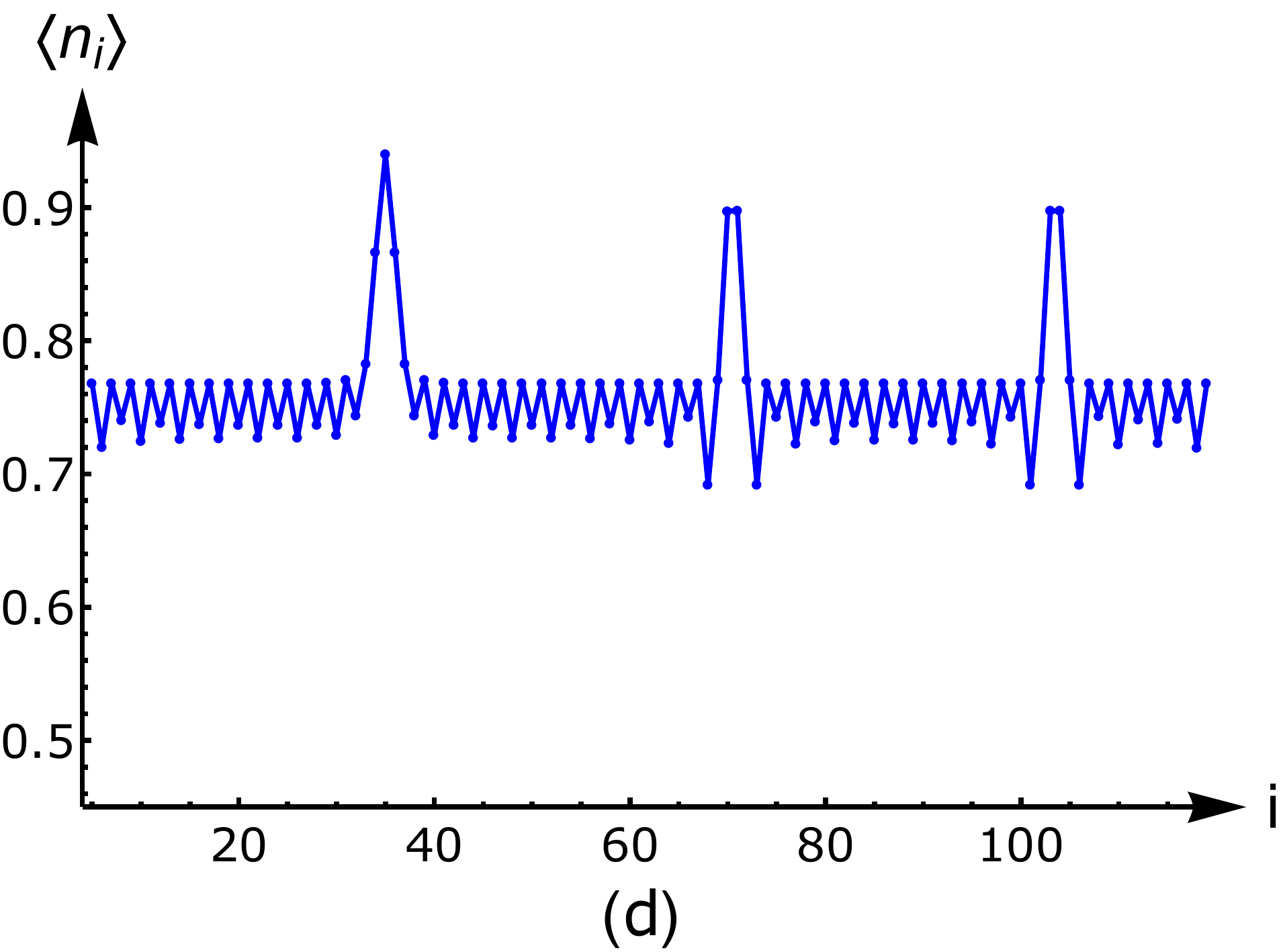}}
\subfigure{\includegraphics[width=4cm]{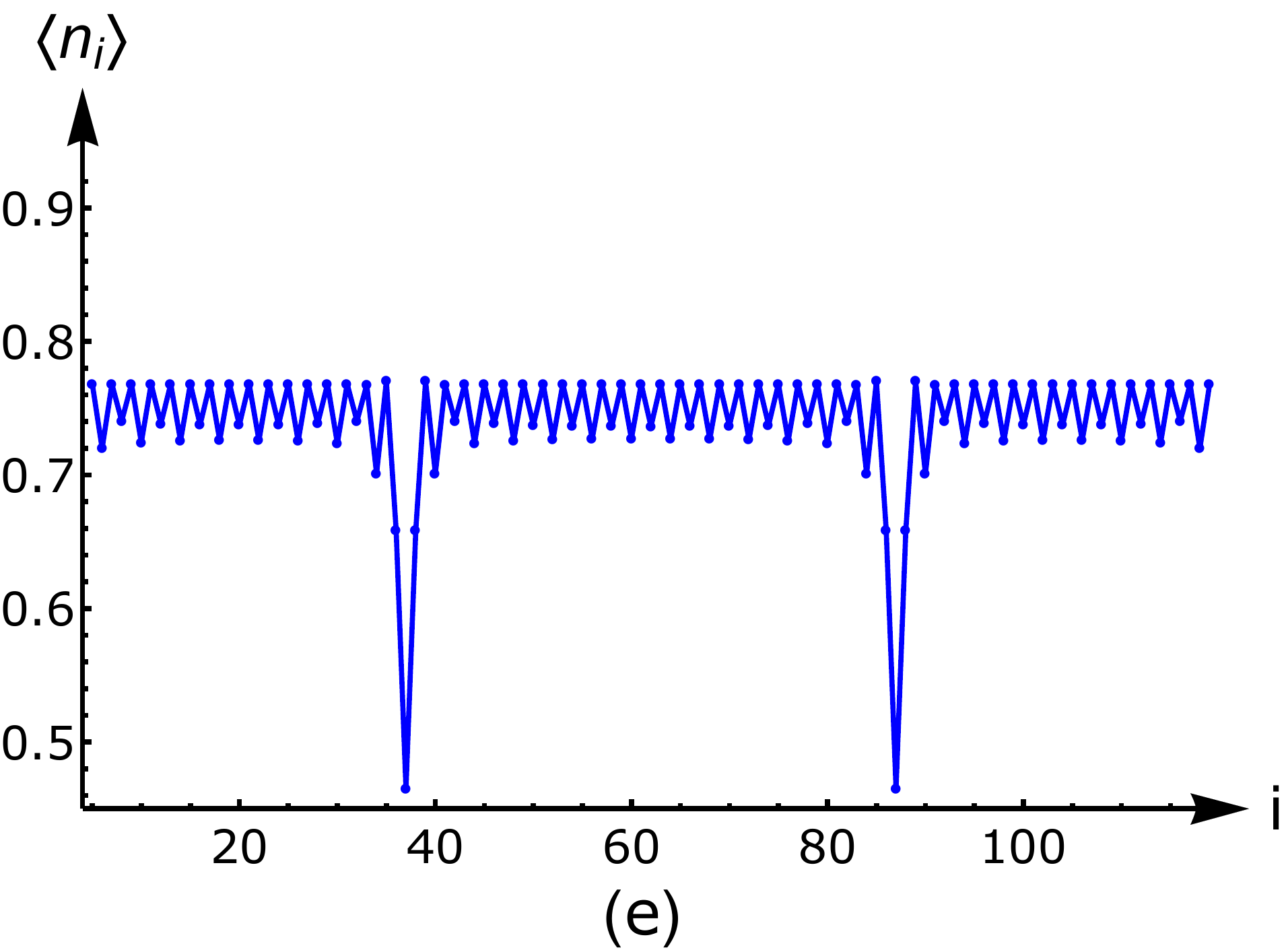}}
\subfigure{\includegraphics[width=4cm]{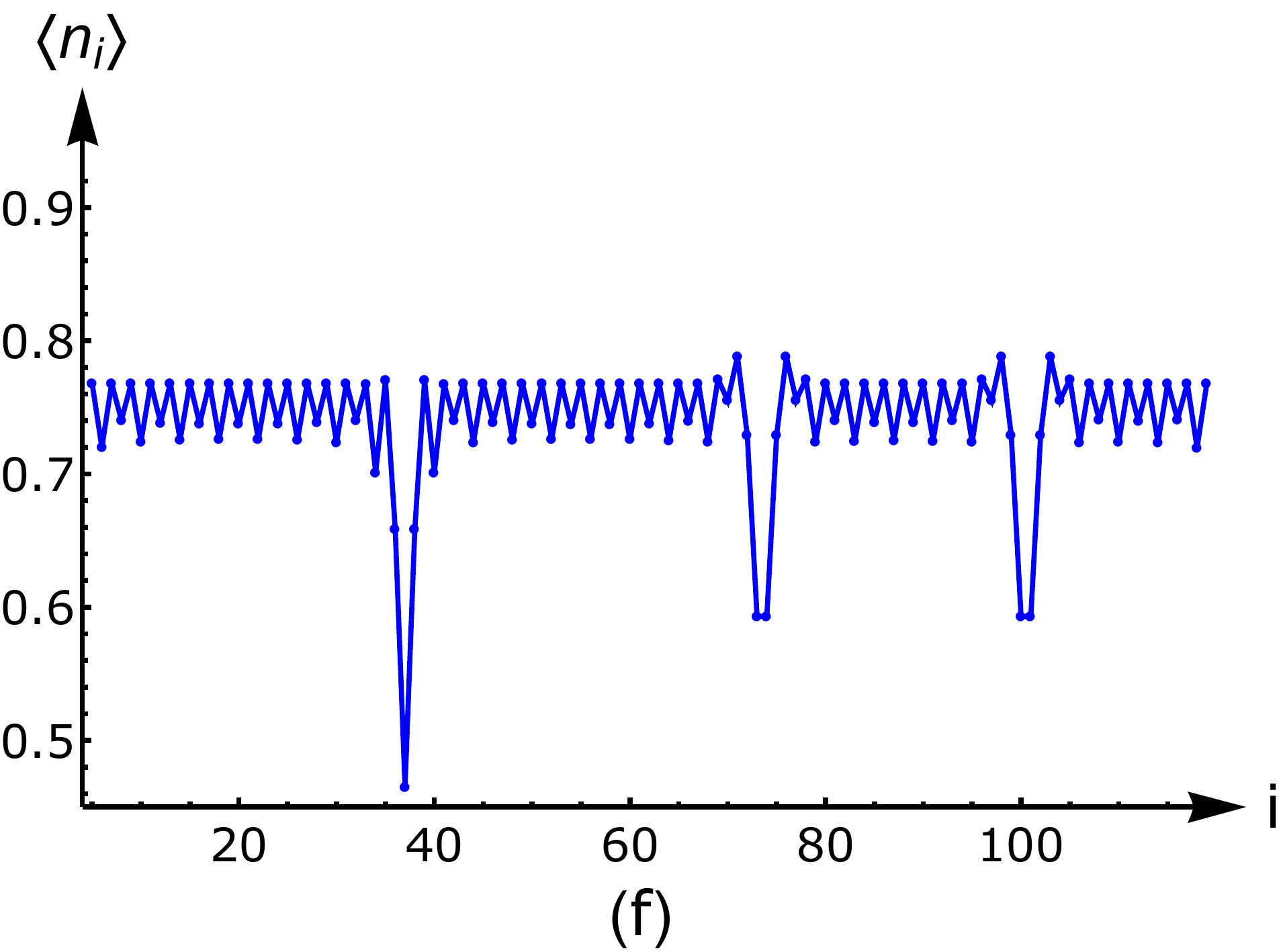}}
\caption{Density profiles of the $L=123$ lattice with $5 \leq i \leq 118$
(excluding  boundary regions).
(a) The ground state with $2\times 31$ doped holes and $S_z^{tot}=1$ supports two neutral solitons. The density and spin are shown in blue and red respectively.
(b) A metastable state with $S_z^{tot}=0$ and $2\times 31$ doped holes.
(c-d) Two metastable states with $S_z^{tot}=0$ and $2\times30$ doped holes.
(e-f) Two metastable states with $S_z^{tot}=0$ and $2\times 32$ doped holes.
}
\label{soliton}
\end{figure}

(1) The ground state density and spin profile in the sector of $S^z_{tot}=1$ are shown in \Fig{soliton}(a). The changes relative to the ground-state  are spread out.  However, it is apparent that the spin-density is doubly peaked, with spin 1/2 in each half of the system, consistent with  the existence of two delocalized spin 1/2 particles.  Manifestly, these particles are neutral.  Moreover, comparing CDW pattern in the middle and  at the boundaries of the cylinder, we find a $\pi$ phase shift.
We conclude that the spin 1 ground-state consists of two delocalized neutral spin-$\frac{1}{2}$ solitons with  $\Delta\alpha_1=\pi$, which we label as $S_0^{1/2}$ in Fig. \ref{wall}(c). The soliton creation energy, $\Delta_{c=0}^{s=1/2}$, is expected to approach half of the spin gap, $\Delta_s$ in the limit $L\to \infty$. As we will see, $\Delta_s$ is sufficiently small, $\Delta_s \lesssim 10^{-4}$, that we cannot determine its $L\to \infty$ value from even the largest system sizes we have studied. The dynamical mass $M^*$ refers to the zero-point energy to confine a soliton to a region of size $L$ according to $E \sim \frac{1}{2 M^*}(\frac{\pi}{L})^2$. As explained in the Supplemental Material, we extract the dynamical mass of spin-$\frac{1}{2}$ soliton $M^*_s\sim 10^{1}$ in the small $L$ region.

(2) A metastable excited state with $S^z_{tot}=0$ can be prepared by applying a proper training field in the initial DMRG simulation, with the result shown in \Fig{soliton}(b).
It contains charge $\pm e$ and spin-0 solitons ($S_{\pm e}^0$) with $\Delta\alpha_1=\pi$.
The solitons are sufficiently ``heavy'' that they remain localized for as many DMRG iterations as we can execute, which also means the dynamical mass of charged solitons $M^*_e$ is effectively infinity. The creation energy of a pair of charge $\pm e$ solitons is $\Delta_{c=e}^{s=0}+\Delta_{c=-e}^{s=0}=0.453$, which is much larger than $\Delta_{c=0}^{s=1/2}$. Note there is no particle-hole symmetry relating the solitons with opposite charge.

(3) The addition of two electrons with $S^z_{tot}=0$ to the ``undoped'' system (with $x=1/4$) results in various configurations, depending on the form of the initial training fields.
In \Fig{soliton}(c), two  $S_e^0$ solitons identical to the left soliton in \Fig{soliton}(b) are clearly seen.
In \Fig{soliton}(d), the right soliton has been broken into two $S_{ e/2}^0$ solitons, each associated with a $\Delta\alpha_1=\pi/2$ domain wall (Fig. \ref{wall}(d)).
By comparing the energies of the states in \Fig{soliton}(c) and \Fig{soliton}(d), we obtain $2\Delta_{c=e/2}^{s=0}-\Delta_{c=e}^{s=0}=0.021$. Similarly, by adding two holes we can obtain the soliton configurations shown in \Fig{soliton}(e) and \Fig{soliton}(f).
In \Fig{soliton}(f), there are two charge $-e/2$ solitons associated with the $\Delta\alpha_1=3\pi/2$ domain walls. By comparing energies in \Fig{soliton}(e) and \Fig{soliton}(f), we obtain $2\Delta_{c=-e/2}^{s=0}-\Delta_{c=-e}^{s=0}=0.350$. A charge $-e$ soliton has much lower creation energy than a pair of $-e/2$ solitons.

\begin{figure}[t]
\centering
\subfigure[]{\label{}\includegraphics[width=4cm]{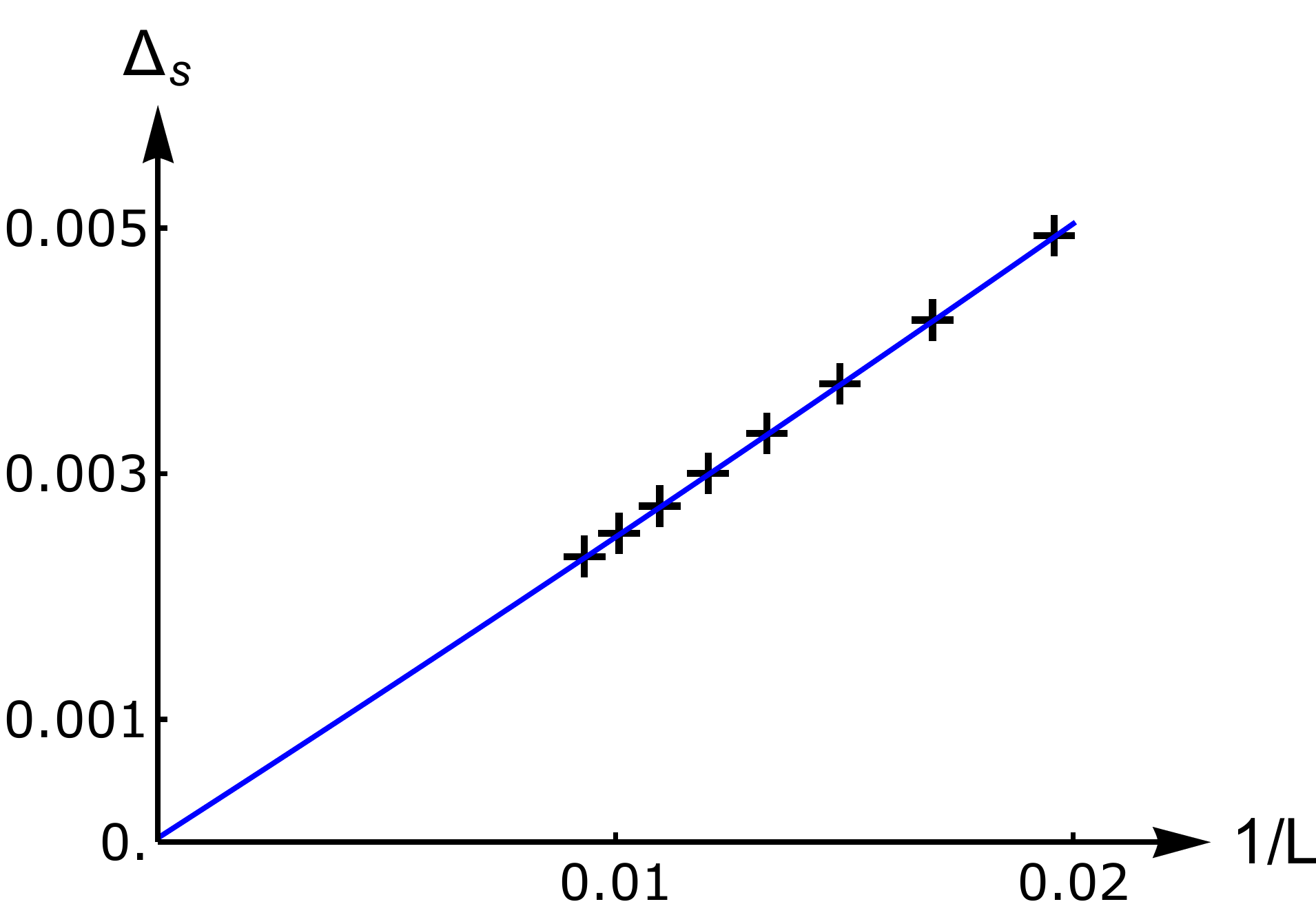}}
\subfigure[]{\label{}\includegraphics[width=4cm]{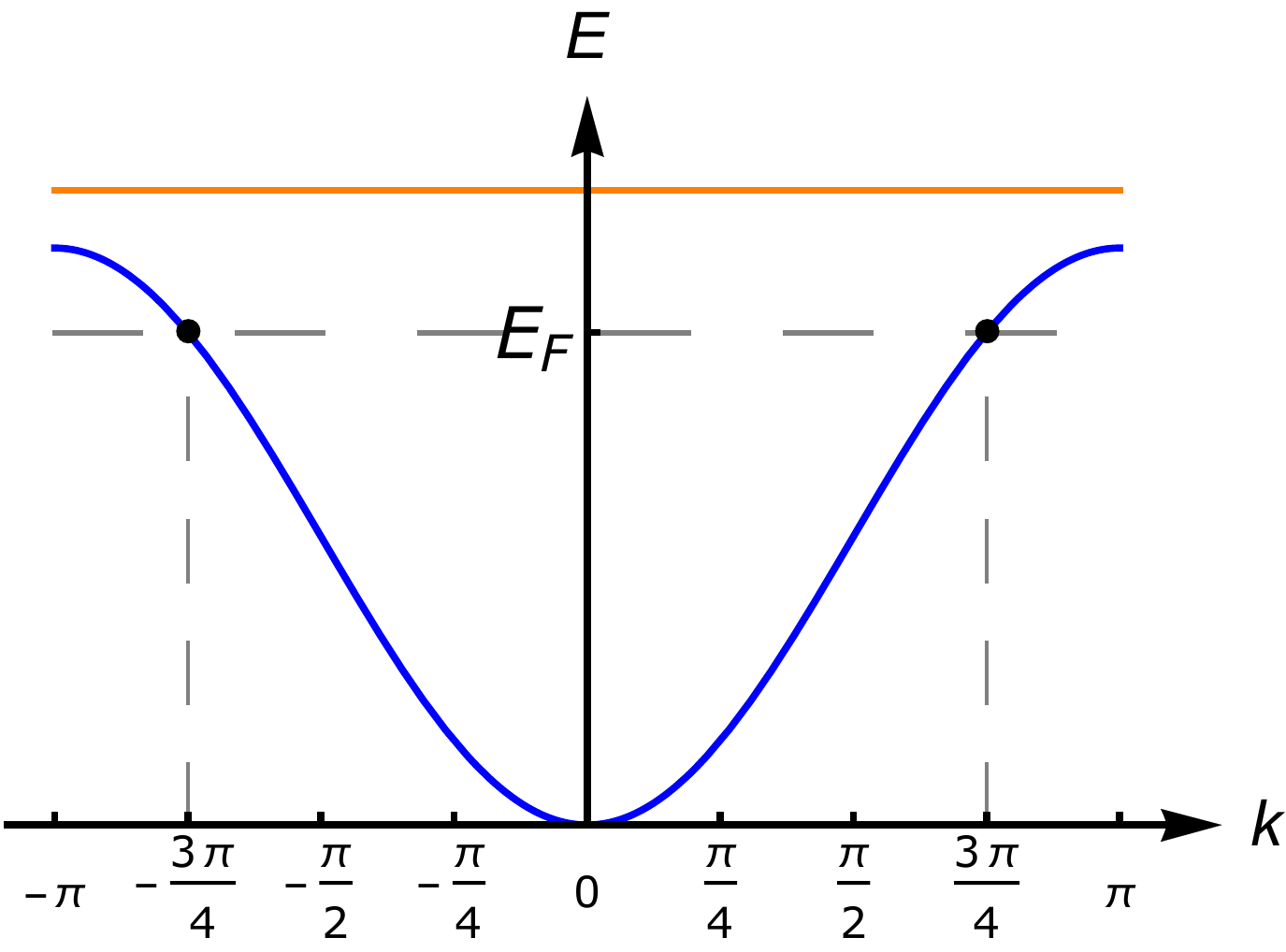}}
\caption{
(a) Dependence of the spin gap on $1/L$; the extrapolation $L\to \infty$ yields $\Delta_s =3.0(1)\times 10^{-5}$. L varies from 51 to 99. (b) The schematic band dispersion of diagonal two-leg cylinder at one quarter doping. The orange line is the unoccupied flat band. The blue one is the dispersive band with two Fermi points (black dots) at $k_F=\pm 3\pi/4$.
}
\label{gap75}
\end{figure}

{\bf Spinon excitation:}
As mentioned above, the spin gap at $x=1/4$ doping is extremely small, which is a novel feature worth further understanding. Because of the period 4 CDW ordering, the enlarged unit cell now has 8 sites and consequently 6 electrons. Thus, consistent with Haldane's conjecture, we should expect a finite spin gap. For finite $L$, $\Delta_s$ is always larger than $0$, but by extrapolation we would infer that $\Delta_s \to 3.0\times 10^{-5}$ as $L\to \infty$, as shown in \Fig{gap75}(a).  This is a small enough value that it could be consistent with $\Delta_s \to 0$.
More importantly, it would imply a spin-correlation length, $\xi_s \sim J/\Delta_s$, which is larger than any accessible system size, making the quantitative aspect of this estimate unreliable. At an intuitive level, the small gap is related to the small value of the principle harmonic of the CDW;  in the limit $\delta \rho_{B\tilde B}\to 0$, the CDW has period 2 with 3 electrons per unit cell, and hence (presumably) no spin-gap.

To flesh out this intuition, we consider the same problem in the context of a ``bosonized'' effective field theory. The non-interacting band structure consists of a flat band and a dispersing band, as shown in \Fig{gap75}(b).
For $x=1/4$, the lower dispersive band is partially filled with $k_F=3 \pi/4$, while the flat band is empty.  Thus, by adiabatic continuity, we expect that the low energy fermionic modes can be expressed in terms of two bosonic fields $\phi_c, \phi_s$ and their duals, $\theta_c, \theta_s$:
\bea
\psi_{\sigma,\lambda}(x) = {\cal N}_{\sigma}e^{i\lambda k_F x}\exp[-i\sqrt{\frac{\pi}{2}} (\theta_{c}+\sigma\theta_{s} +\lambda\phi_{c} + \lambda\sigma\phi_{s})],\nn
\eea
where $\sigma=\pm 1$ is the polarization of the spin and $\lambda=\pm 1$ for right and left moving fermions. The  period 2 and 4 CDW orders come from the expectation value of ${\cal O}_{4k_F} \equiv \psi_{\A,+}^\dagger \psi_{\V,+}^\dagger\psi_{ \V,-} \psi_{ \A,-}$ and ${\cal O}_{2k_F} \equiv \psi_{\sigma,+}^\dagger\psi_{ \sigma,-}$ respectively:
\bea
{\cal O}_{4k_F} &&= {\cal N}_{4k_F}e^{i3\pi x}e^{i\sqrt{8\pi}\phi_{c}},\\
{\cal O}_{2k_F} &&= {\cal N}_{2k_F}e^{i3\pi x/2}e^{i\sqrt{2\pi}\phi_{c}}\cos[\sqrt{2\pi}\phi_{s}].\label{eq5}
\eea
Because of the $\cos[\sqrt{2\pi}\phi_{s}]$ factor in \Eq{eq5}, ordering of ${\cal O}_{2k_F}$, {\it i.e.} a period 4 CDW, requires condensing $\phi_s$, which gives rise to a finite spin gap $\Delta_{s}$.
To obtain an estimate of the expected gap magnitude, we invoke the expected scaling relations $\langle e^{i\sqrt{8\pi}\phi_{c}}\rangle \sim [\langle e^{i\sqrt{2\pi}\phi_{c}}\rangle]^4$ and $\langle\cos[\sqrt{2\pi}\phi_{s}] \rangle \sim \sqrt{\Delta_s/\Omega} $ where $\Omega$ is a UV cutoff to obtain
\bea
\Delta_s\sim \frac{\avg{{\cal O}_{2k_F}}^2}{\sqrt{\avg{{\cal O}_{4k_F}}}}\Omega.
\label{spingap}
\eea
By further identifying $\avg{{\cal O}_{2k_F}}\sim \delta\rho_{\tilde B B}$, $\avg{{\cal O}_{4k_F}}\sim \delta\rho_{A\bar B}$, and $\Omega \sim t$,
we estimate $\Delta_s\sim 4 \times 10^{-5}$ which is small and remarkably consistent with the estimate obtained from finite-size scaling.

{\bf Concluding remarks:} From both numerical results and bosonization analysis, we infer that the creation energy of the spinon $\Delta_{c=0}^{s=1/2}$ is extremely small. This is a quite surprising result; the creation energies of the charged solitons are three or four orders of magnitude larger. Without any fine tuning or small parameters in the microscopic model, a striking mass hierarchy emerges in the low energy physics of the $t$-$J$ model on the diagonal two-leg cylinder!

We have also carried out similar DMRG studies for values of  $J/t$ other than 1/3, including $J/t=1/4,1/6,1/10$.  We find qualitatively similar results both for the fractional quantum numbers of the solitons at $x=1/4$ and the mass hierarchy. For other values of $x$, still more complicated forms of solitons arise.
A systematic study of the phase diagram as a function of both $J/t$ and $x$ will be discussed in future work.

{\it Acknowledgement}: YFJ and HY were supported in part by the NSFC under Grant No. 11474175 at Tsinghua University. HCJ was supported by the Department of Energy, Office of Science, Basic Energy Sciences, Materials Sciences and Engineering Division, under Contract DE-AC02-76SF00515. SAK was supported in part by NSF grant $\#$DMR 1265593 at Stanford University.

\section{SUPPLEMENTAL MATERIALS}
\renewcommand{\theequation}{S\arabic{equation}}
\setcounter{equation}{0}
\renewcommand{\thefigure}{S\arabic{figure}}
\setcounter{figure}{0}
\renewcommand{\thetable}{S\arabic{table}}
\setcounter{table}{0}
\subsection{A. Details of DMRG simulations}
Due to the delicate metastable states, the DMRG simulation of diagonal cylinder easily converges at a local minimum. To encounter this problem, we apply a training field in the initial step of DMRG simulation. In our model, the simple external training field term in Hamiltonian reads:
\bea
H_{train}&=&\sum_{x,y,\s}u(x,y)n_{x,y,\s} \nn\\
u(x,y)&=& u_0(-1)^x \max(0,\frac{N_0-n_{sweep}}{N_0})\ , \label{train}
\eea
here $u_0\sim 10^0$ is a constant number, and $n_{sweep}$ counts the DMRG sweep. The external potential $u(x,y)$ plays a role as a training field which is gradually reduced during DMRG sweeps. In our calculation, this training term is finally removed after 14 sweeps ($N_0=15$).

The initial training field in \Eq{train} leads to a perfect CDW state which has the lowest energy. More importantly, via $H_{train}$ we can even take advantage of those high energy metastable states to study the property of soliton excitations. By slightly changing the form of $u(x,y)$, we can create different CDW domain walls and study the physics property of soliton excitations associated with them.

\subsection{B. Creation energy and dynamical mass of solitons}
We study two types of effective masses of the solitons. One is the creation energy $\Delta$, which in the context of a relativistic quantum field theory is referred to as the mass. The second is the dynamical mass $M^*$, which determines the extent to which the soliton tends to delocalized - specifically, the energy to localize the soliton in a (large) box of length $L$ is $\frac{1}{2 M^*}(\frac{\pi}{L})^2 $.

For charged solitons, we can measure their creation energy by creating a pair of solitons with opposite charges. However, their dynamical masses are almost infinite within the present level of computational accuracy because they remain localized even after hundreds of DMRG sweeps.

For spin-1/2 solitons, both their creation energy and dynamical masses are much smaller than the charged ones. We can measure the dynamical mass of the spin solitons by looking at the energy of two soliton states as a function of system size. As an extended excitation, the interaction between the two solitons need be considered. For small enough $L$, we can write down a perturbative theory in powers of interaction $V$: 
\bea
\Delta_s(L)&=&\Delta_s(\infty)+\frac{1}{2 M^*}\frac{5\pi^2}{L^2}+\frac{V}{L}\big[A+B L^{-2}+\cdots\big] \nn\\
&& -V^2 M^* L^{-1}\big[C+\cdots\big]+\cdots
\eea
where $\Delta_s(\infty)$ is the creation energy, $A$, $B$ and $C$ are constant. The perturbation theory breaks down at large $L$. By fitting $\Delta_s(L)$ at small $L$ region, we obtain $M^*\approx 28\sim O(10^1)$.

\begin{figure}[t]
\centering
\includegraphics[width=7cm]{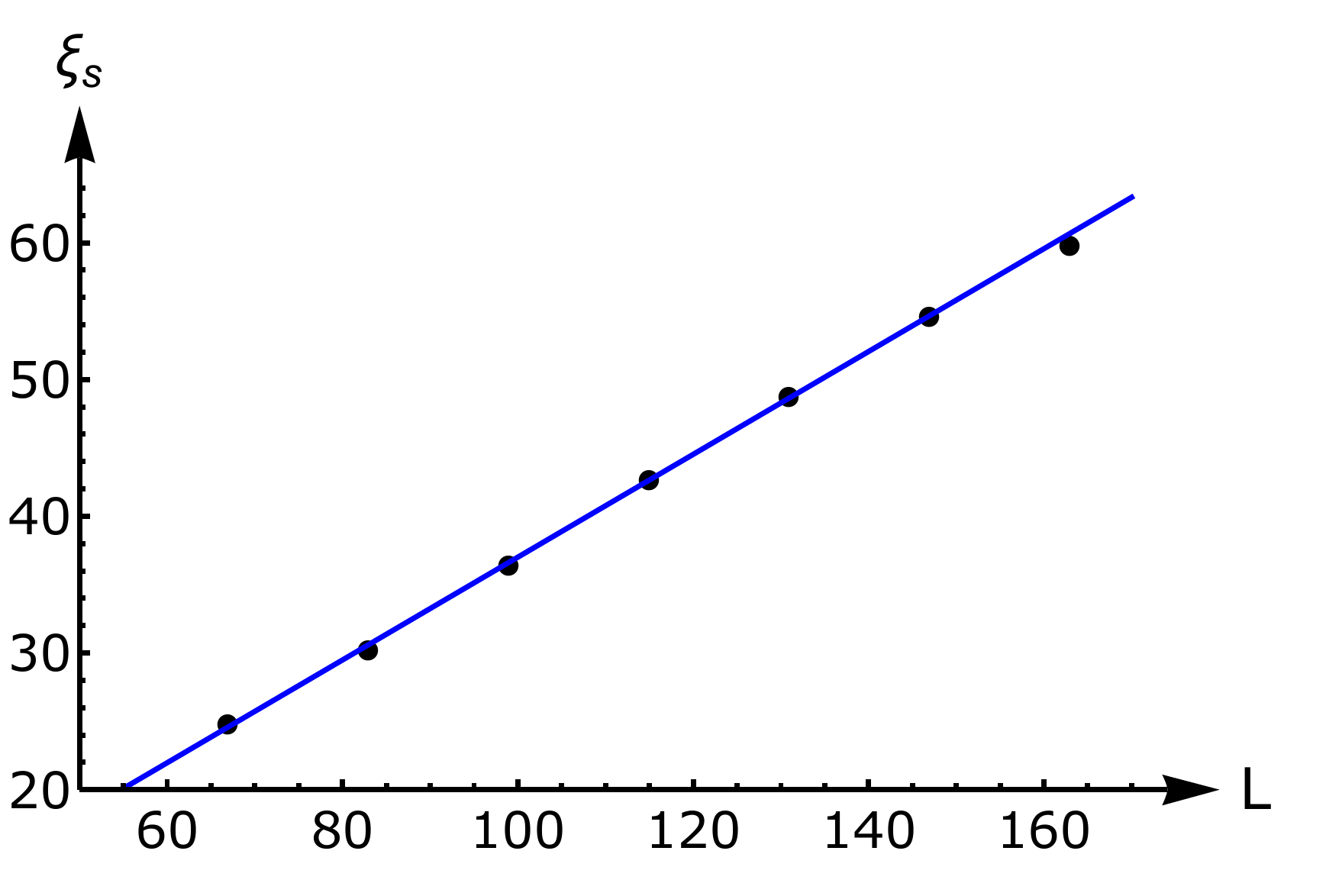}
\caption{Length dependence of the spin-spin correlation length $\xi_s$ on system varied from $L=67$ to $163$.
}
\label{corS75}
\end{figure}

\subsection{C. Spin-spin correlations}
On the relatively small systems (comparing with large $\xi_s \sim J/\Delta_s$), we measure the spin-spin correlation functions in a higher accuracy $\e_{trun} <10^{-10}$ and find a rather long correlation length which is compatible to the system size. As the cylinder size becomes longer, the correlation length increases, as shown in \Fig{corS75}, which supports that the correlation length is restricted by finite system size for the systems we have studied. The longest $L=163$ data shows a signature of saturation which weakly implies a finite correlation length in the thermodynamic limit.

\end{document}